\documentclass{article}

\usepackage[preprint, nonatbib]{neurips_2019}

\usepackage[utf8]{inputenc} 
\usepackage[T1]{fontenc}    
\usepackage{hyperref}       
\usepackage{url}            
\usepackage{booktabs}       
\usepackage{amsfonts}       
\usepackage{nicefrac}       
\usepackage{microtype}      
\usepackage{graphicx}

\title{Development and Implementation of a Dashboard for Diabetes Care Management in OpenMRS}

\author{%
  Bhanu Teja Yandrapalli\thanks{This is a shortened version of the Capstone Project that was accepted by the Faculty of Indiana University, in partial fulfillment of the requirements for the degree of Master of Science in Health Informatics.}\\
  Department of BioHealth Informatics\\
  Indiana University-Purdue University\\
  Indianapolis, IN 46202 \\
  \texttt{byandrap@iu.edu}\\
  \And
  Josette Jones\\
  Department of BioHealth Informatics\\
  Indiana University-Purdue University\\
  Indianapolis, IN 46202 \\
  \texttt{jofjones@iupui.edu} \\
  \And
  Saptarshi Purkayastha\\
  Department of BioHealth Informatics\\
  Indiana University-Purdue University\\
  Indianapolis, IN 46202 \\
  \texttt{saptpurk@iupui.edu} \\
 }

\begin{document}
\maketitle
\begin{abstract}
   A clinical dashboard for a patient’s diabetes condition helps physicians to make better decisions based on readily available information. OpenMRS is a widely used open-source electronic health records system but does not provide a disease-specific dashboard. This project implemented a dashboard for displaying all diabetes-related lab measures at one place, when a physician accesses a patient record in OpenMRS. It summarizes a list of diabetes-related clinical measures through an intuitive, chart-based, customizable user experience. Gauge charts are used to display the most important lab values for Glucose, Renal Function, and Lipid Profile tests. Data Tables are used to display data of the lab values from the past and current visit in the table, including the ability to search for a specific visit date. Interactive line charts are used to display the trends of lab measures. Diabetes Dashboard may help physicians to make quicker decisions through this snapshot view. We took data for few patients and demonstrated this to clinicians as a proof of concept, without performing a full-fledged user evaluation. Future work involves integrating this dashboard with clinical practice guidelines and alerting when measures are outside the guidelines.


\end{abstract}

\section{Introduction}
Diabetes Mellitus is a chronic metabolic disorder which is characterized by a persistent increase in the blood glucose level. It is caused due to the deficiency in the production of insulin by the pancreas or ineffectiveness of insulin produced by the body. Diabetes mellitus is categorized into Type 1 Diabetes Mellitus, Type 2 Diabetes mellitus and other types of diabetes. Diabetes is diagnosed by doing an HbA1c test, Fasting Plasma Glucose test (FPGT), and Oral Glucose tolerance test (OGTT) \cite{association_2._2018}.

Social acceptance for the use of Information Technology in health care and the legislation for incentivizing the use of electronic health/medical records (EHR/EMR) has led to additional burden on physicians for storing a lot of documentation in these systems \cite{mennemeyer_impact_2016}. EHR is an electronic version of the individual's health care record, which can be accessed by a physician in real-time. EHR systems today store all kinds of information such as medical history, diagnoses, medications, treatment plans, allergy information, radiology images, laboratory results, etc. However, physicians have often complained about the lack of evidence-based tools that can help them in improving their decision-making speed and accuracy in the EHR systems \cite{zimlichman_return_2013}. 

OpenMRS is an EHR system, collaboratively developed by a large number of volunteers spread across the world \cite{mamlin_cooking_2006}. OpenMRS is also a software platform and reference application which enables the design of customized medical records, which several researchers have used to prototype upcoming health IT standards \cite{gichoya_platform_2018}. OpenMRS is also among the most simplest EHRs in terms of user-performance \cite{purkayastha_comparison_2019}. The system is built on a conceptual database structure with no dependence on the medical information's actual types. Thus, making it flexible for use for a number of different diseases and purposes. There is a concept dictionary which stores all the knowledge base in the form of terms against which the actual values are stored. OpenMRS also provides user-authentication, privilege-based access which allows different users to have differential access to the system \cite{purkayastha_implementation_2017}. The patient repository features include the creation of patient data, maintenance, demographics, encounter, orders, data and clinical observations (e.g., like lab results). A recent innovation in OpenMRS is the use of external web applications through the Open Web Apps Module, based on the W3C Open Web App standard \cite{puder_exposing_2014}.

Decision science researchers have suggested that dashboards provide an overview of information, which might improve decision making \cite{franklin_dashboard_2017}. In health care, such dashboards can provide progression of patient disease, with infographics that make it easy to process complex and disparate information \cite{franklin_dashboard_2017}. The graphical representation of data also helps the physician to assess the results faster instead of going through each data point. Dashboards may also incorporate statistical visualization using trends of lab values, which saves time for the physician in identifying the outliers easily and correlations of data \cite{brasoveanu_visualizing_2017}.

In this paper, we describe the implementation of a Diabetes Dashboard in OpenMRS, which is a web-based tool to provide summarized information about patient blood glucose profile, lipid and renal profile of the last visit and a data table to show the changes in the values of lab parameters over time and graphical interface shows the trends of lab values. The information makes the physician's work simple, since with a click the physician will be able to view patient lab results and discover disease trajectory at a glance. Dashboards save time for physicians and reduce going through multiple pages of visits to access the patient's lab data. With this dashboard, the physician has access to all the essential information for diabetes management with a single click.

\section{Background}
Diabetes Dashboard is an open web app installed on OpenMRS. Different ideas on how to develop the dashboard were gathered from existing literature. The primary idea for this Diabetes Dashboard is taken from the article written by Sim et al. \cite{sim_development_2017}. The authors have done a great job in designing a dashboard and integrating with the existing health record system at a University hospital. The overview of our app and concept is adapted from the Sim et al. article. The data elements such as LDL, HbA1c, and other elements are used for the assessment in this dashboard are standard ones \cite{koopman_diabetes_2011, dowding_dashboards_2015}. The idea of developing a dashboard with color-coding to identify the targets which can give a clear picture of the status of the patient's disease progression was taken from the article written by Martinez et al. \cite{martinez_patient-facing_2018}. Their study provides clarity that clinical measures such as HbA1c, LDL-C were to be considered and gauge charts are to be used to represent important lab results. The idea of lab measures displayed over time was taken from the article written by Stacy Zahanova et al. \cite{zahanova_iscreen_2017}. Traffic light approach with color-coding is taken from Fonda et al. \cite{fonda_usability_2008} and Dagliati et al.\cite{dagliati_dashboard-based_2018}.

\section{Methodology}
We used the following MeSH Terms to search for literature related to previous research done on dashboards for diabetes: Diabetes, Diabetes Dashboard, Clinical Decision Support System (CDSS), Electronic Health Record (EHR)/Electronic Medical Record (EMR), Glycemic Control, Open Web App (OWA), HbA1c, Clinical variables, Diabetic measures.

\textbf{Search Resources:} For the literature search we used resources such as PubMed, Science Direct, and Ovid. Mendeley was used as the citation manager.

\textbf{Search Strategy:} Only research articles in the English language were included in the study. Since there were not many results, no search restrictions were included for this project.

\subsection{Methods}
OpenMRS customization is usually done using Java modules. However, recent architectural changes such as an elaborate REST API has made it feasible to use Open Web Apps (HTML, CSS, JavaScript apps), by loading them using the OpenWebApp (OWA) module. As mentioned earlier, OpenMRS lacks the functionality to display all the vital data of a diabetic individual in one place. So, we developed a proof-of-concept OWA-based Diabetes Dashboard. Data about diabetes lab observations was imported into a dummy OpenMRS patient. The OWA interface is developed using HTML, CSS, JavaScript. We used high charts (highccharts.com) for JavaScript-based interactive charting. Data over time is shown using data tables from (datatables.net). The lab measure reference ranges are taken from the American Diabetes Association. The following sections provide a detailed overview of HTML form and developed diabetes dashboard.

\subsubsection{Design of HTML Form}
An HTML lab report form is developed by utilizing the HTML form entry module, which will enable the user to create forms. Lab report form created has fields in three sections: Glycemic profile, Lipid profile, Renal Function tests along with other fields to capture the vitals of the patient. The OpenMRS concept dictionary contains all the medical concepts and descriptions. Many common concepts were already present in the dictionary. Concepts which were not present were created using the dictionary module. Medical concepts are mapped to terminologies like LOINC and SNOMED. Each concept has a Universal Unique Identifier (UUID), which is used to retrieve data between EHR system and OWA. The designed form has HbA1c, Fasting Plasma Glucose test (FPGT) in glycemic profile as referenced from various articles from literature. HbA1c is one of the important markers considered as a standard measure to evaluate the progression of diabetes. In the Lipid profile, following are the variables considered: Total cholesterol, Low-Density Lipoproteins (LDL-C), High-Density Lipoproteins (LDL-C), Triglycerides (TG). In the renal function tests, serum creatinine and eGFR are taken. Unlike in the article by Sim et al.\cite{sim_development_2017}, there is no specific laboratory system in place to capture all the details in segregated fashion. After reviewing the literature and American Diabetes Association guidelines for the diagnosis and management following clinical variables are selected and are included in the form.

When a test is ordered for a patient with diabetes by a physician, upon completion of the analysis the results are entered manually into the created HTML form. The lab technician will log in into OpenMRS and find the patient using find options, under the forms section there is a diabetes form which needed to be selected to enter the results in respective fields. To prevent any manual errors there is a validation set for the concepts at the back-end.

\subsubsection{Design and Development of Diabetes Dashboard}
Tools used in Web app development: Node.js, Yeoman OWA OpenMRS generator, HTML, CSS, JavaScript, High Charts, Data tables. The first step is to successfully  create an OWA.
Open Web App (OWA) is a standard packaging format of HTML, CSS, and JavaScript which can be used to build applications that can be run across different browsers and mobile devices. Node.js is a framework for building a Web application, Yeoman OpenMRS OWA generator is used to scaffold the Open web app. Node Package Manager will allow the yeoman script installation which is necessary for building libraries of OWA package. The first step is to make sure Node.js 6+ is installed and open terminal use the command found on (https://www.npmjs.com/package/@openmrs/generator-openmrs-owa) step by step to install dependencies, run OWA generator and Scaffold the Open Web App. It will ask to give the app name, description and libraries to be included in the Open Web App package. The app is successfully scaffolded. REST API services are used to get or fetch the data from the EHR into a web app.

After the OWA package is installed, it is linked as a widget under general actions on the patient record page. The code is hosted on GitHub at (https://github.com/nani147/owa-diabetesdashboard) for any future collaboration. The OWA created is a package of index HTML file where the code for the app goes, an img folder for holding images, CSS folder to hold files essential for styling the HTML document and js Folder where the JavaScript goes to bring functionalities for the app.

\begin{figure}[h]
\centering
\includegraphics[width=0.7\textwidth]{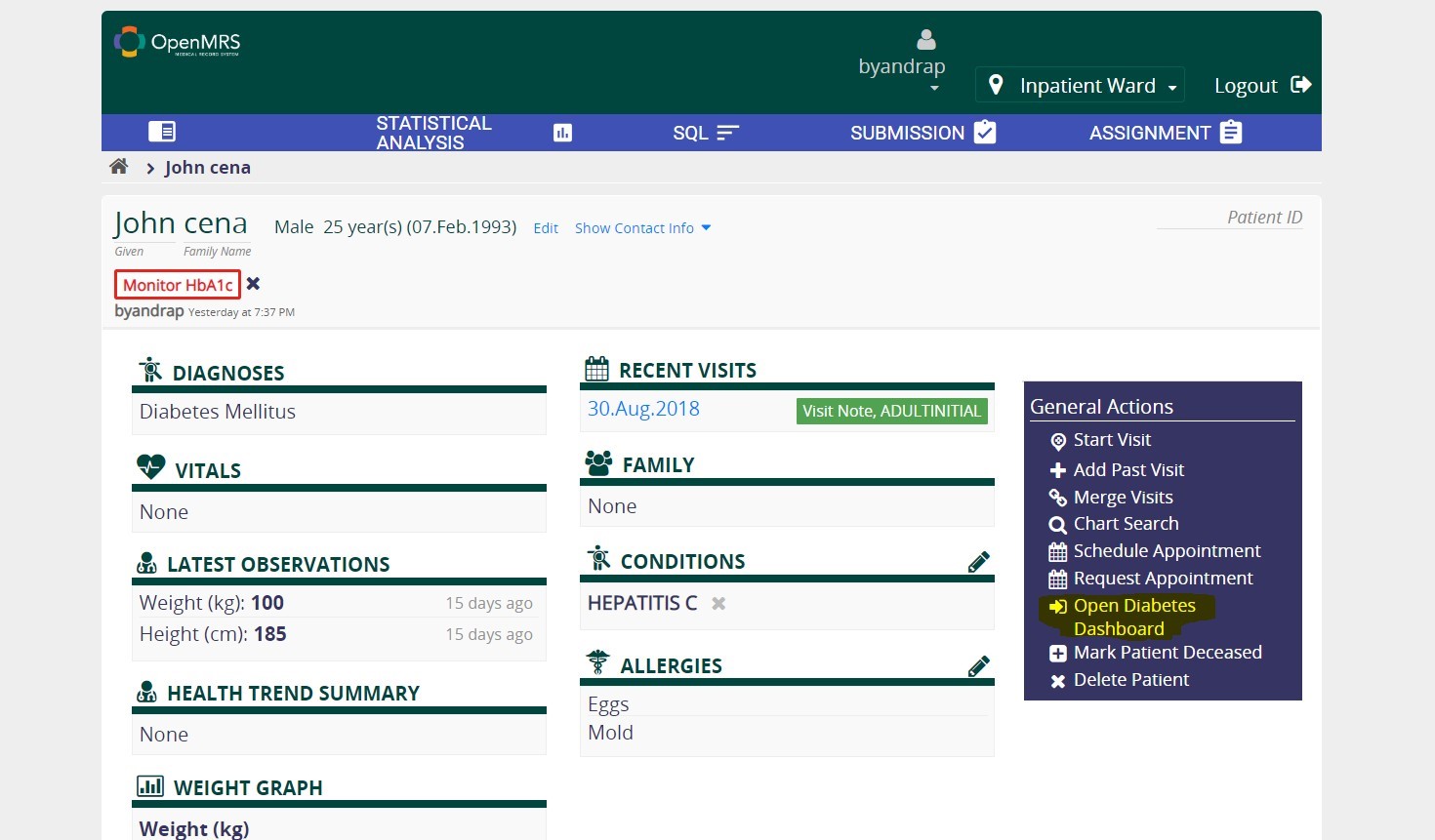}
\caption{Presence of direct link in general actions}
\label{fig:my_label}
\end{figure}

Figure 1 is a screenshot to show how a physician can access the Diabetes dashboard from the OpenMRS Reference Application patient dashboard. The highlighted Open Diabetes dashboard is placed under general actions as a widget that can be easily accessible by physicians. The widget is linked to the dashboard by using the manage apps module in OpenMRS, where a URL link is assigned an icon to link to the app. When physicians click on the link it takes to the dashboard which appears like in Figure 2.

\begin{figure}[h!]
\centering
\includegraphics[width=0.7\textwidth]{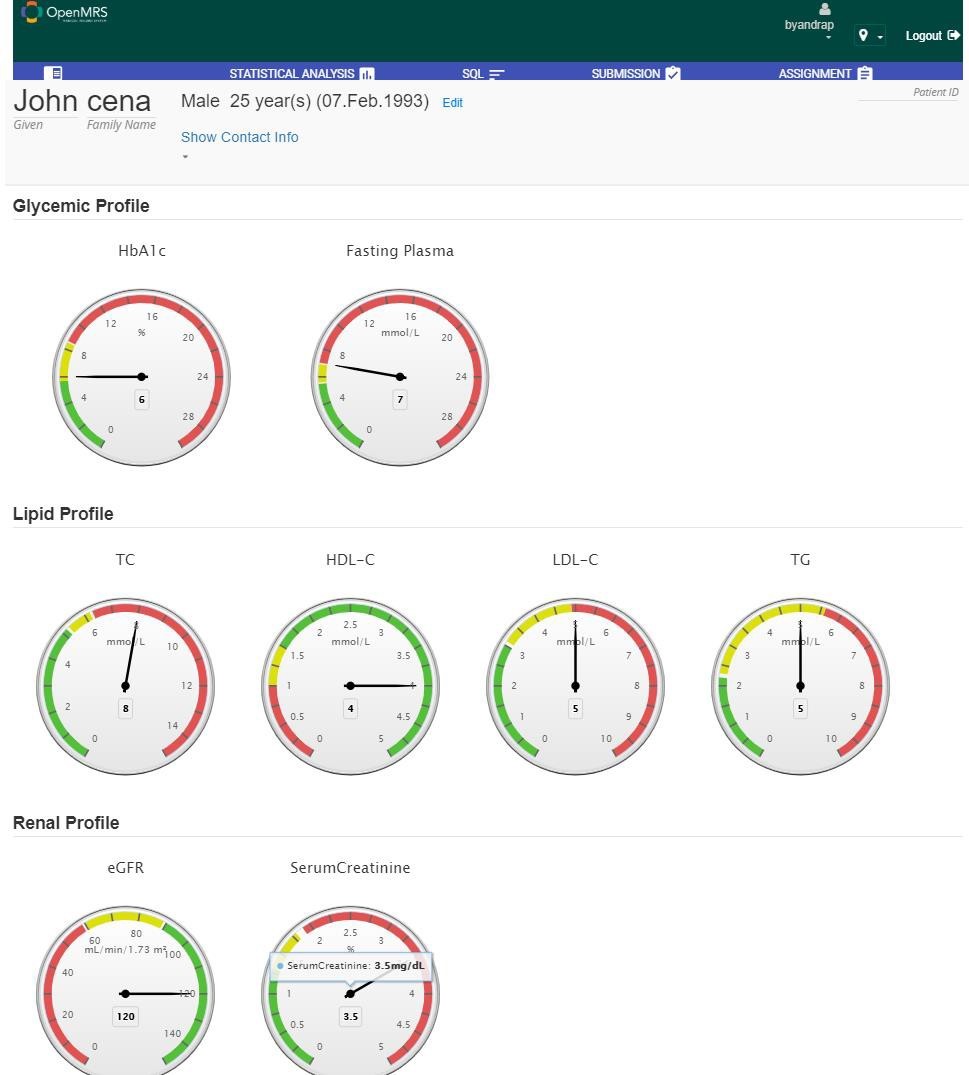}
\caption{Patient header and Gauge Charts with color coding}
\label{fig:my_label}
\end{figure}

Due to the space limitations (the page scrolls vertically), the whole app cannot be shown, but in a single view, the physician can view the visualization, data tables and graphs, as shown in Figure 3.

\begin{figure}[h!]
 \centering
 \includegraphics[width=0.8\textwidth]{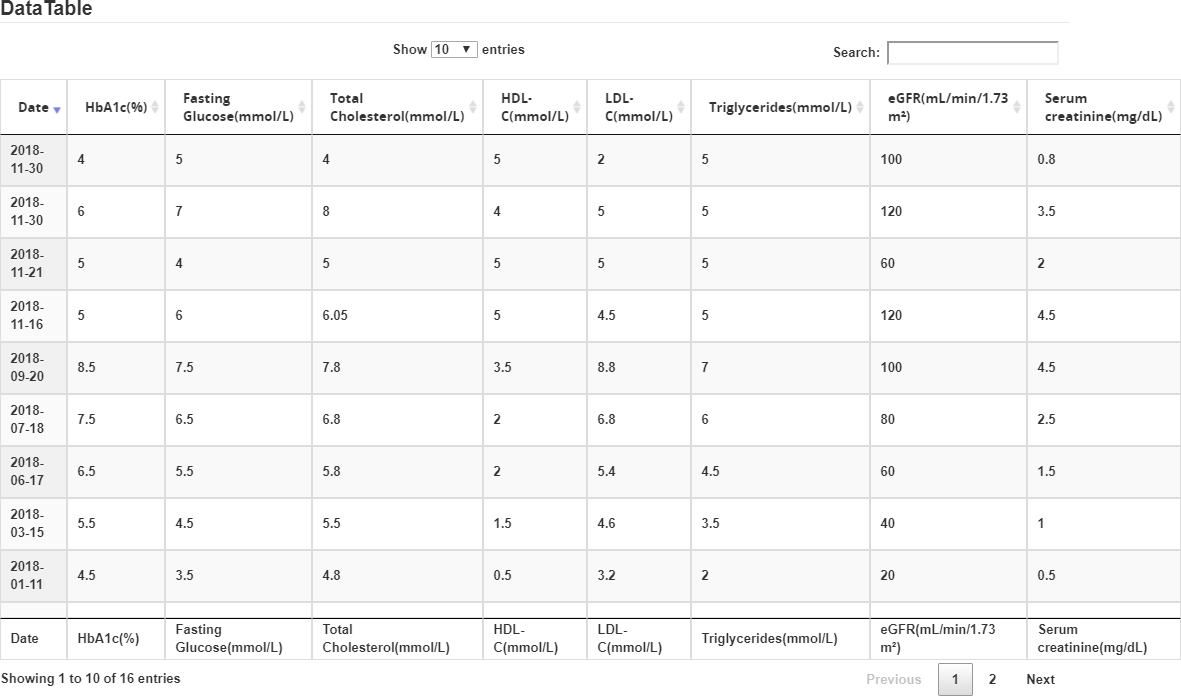}
 \caption{Data table to show observation values over time}
 \label{fig:my_label}
\end{figure}

On the top of the dashboard is the patient header containing the details of the patient and it is displayed using a JavaScript. The patient header details are loaded into the app as a header by utilizing the target \textit{onload} function and the details of the respective patient are displayed.

\textbf{Gauge Charts:}

Circular plots also known as Gauge Charts, are an interactive chart built using JavaScript, with dial pointing to indicate the values of the glycemic, renal and lipid profiles of the patient. We use color coding to highlight normal/abnormal values for these observations, as can be seen in Figure 2. In the HTML for gauges, containers are assigned for each of the lab measures and AJAX JavaScript code gets the values from the \textit{/obs} REST API endpoint and displays the values of the observations into the gauges. The response of the REST API is stored in the browser \textit{localstorage}, so that the app can also be used offline, once the REST calls have been made and the values have not been updated in the EHR. This makes the diabetes dashboard more responsive. The code is written in a way to display the last value, i.e. from the recent visit. In the Gauge chart, we use the color codes according to the “\textit{Traffic light}” approach of red (abnormal), yellow (slightly above normal) and green colors (normal) are used to highlight the lab measures. Plot Band in Gauge Series is used to define the color coding according to the standard measures specified in ADA. For the three profiles, data is retrieved from the EHR system and is displayed at different color regions depending on the lab measure value, if the values are under the preferred range the dial will point towards the values and will be in the green zone and so on for the colored regions in Gauge. We enabled Tooltip to allow hovering over an option, and show the values along with the measure name. These Gauge charts will help physicians to have a detailed overview of diabetes clinical measures. At a glance the physician can see what clinical measures are not in range.

\begin{figure}[h!]
 \centering
 \includegraphics[width=0.7\textwidth]{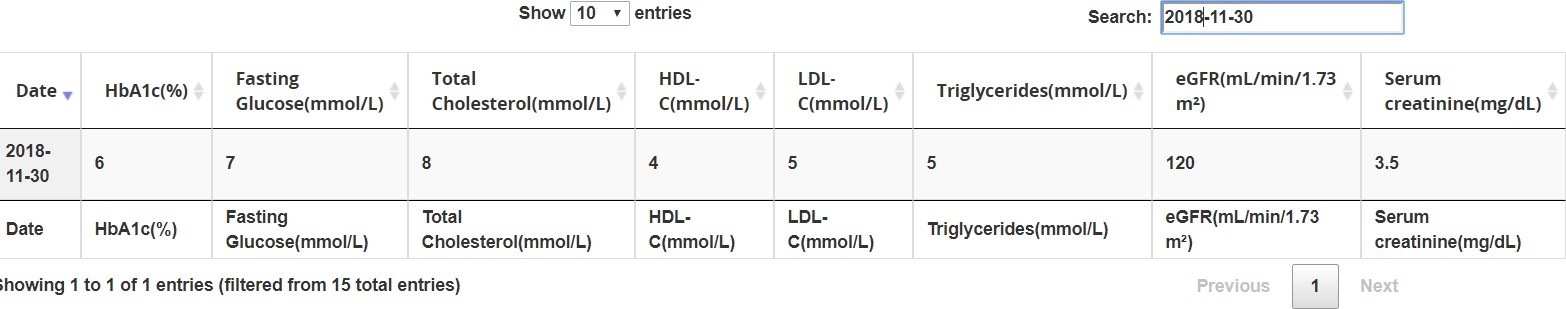}
 \caption{Data table with a date search}
 \label{fig:my_label}
\end{figure}

\textbf{Data Table:}

Data Tables is a plug-in for the jQuery JavaScript library. It is a highly flexible tool, build upon the foundations of progressive enhancement, that adds all these advanced features to any HTML table (datatables.net). Data Table displays the information from the previous visits by utilizing Server-Side-Processing. An AJAX GET request is made to the specified URL in OpenMRS REST API and the data is returned to the table along with respective dates. Our Data Table displays information of the visits according to dates and the records can be limited to 10 to 100 by using the show icon. The table also has a search bar which can be used to search the visit date of the patient. Figure 5 shows the data is searched for 2018-11-30, which displays the clinical measures for the specified date. This will help the physician to see for particular visit date data.The summary data which is displayed in gauge charts are only for the recent visit, but data tables can do more than that. The data table is useful because it can show the data over time, which includes the past visit information on the patient’s performance. With this feature doctor can see the measures for the past and recent Visit and compare both values that can help to identify the measures outliers, from this physician will have an overview of data and can talk to the patient about any interventions or treatment change.

\begin{figure}[h!]
 \centering
 \includegraphics[width=0.3\textwidth]{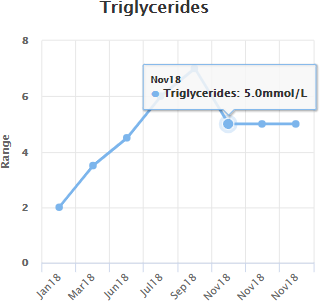}
 \caption{A Triglycerides line chart showing trend over time}
 \label{fig:my_label}
\end{figure}

\textbf{Lab Parameter Trends:}

To show the trends of lab values, High chart Line graphs are used to display. These are interactive line-graphs where physicians can hover over the data points to see the value of the lab measure and month which the test is done. To get all the values of the previous visits, an Ajax call is done for each UUID and data is pushed into an array and “local storage” property is used to store data for easy retrieval of data, code is written in line chart Js file to the get the lab measures values into the array, also to retrieve date of the test. On the x-axis, months of the test done is displayed and on the y-axis is range. These line graphs will help physicians to see the trends of lab measures, serves as a better visualization. Like in the below snippet physician can interact with the line graphs to see the values of different lab measures.

\begin{figure}[h!]
 \centering
 \includegraphics[width=0.6\textwidth]{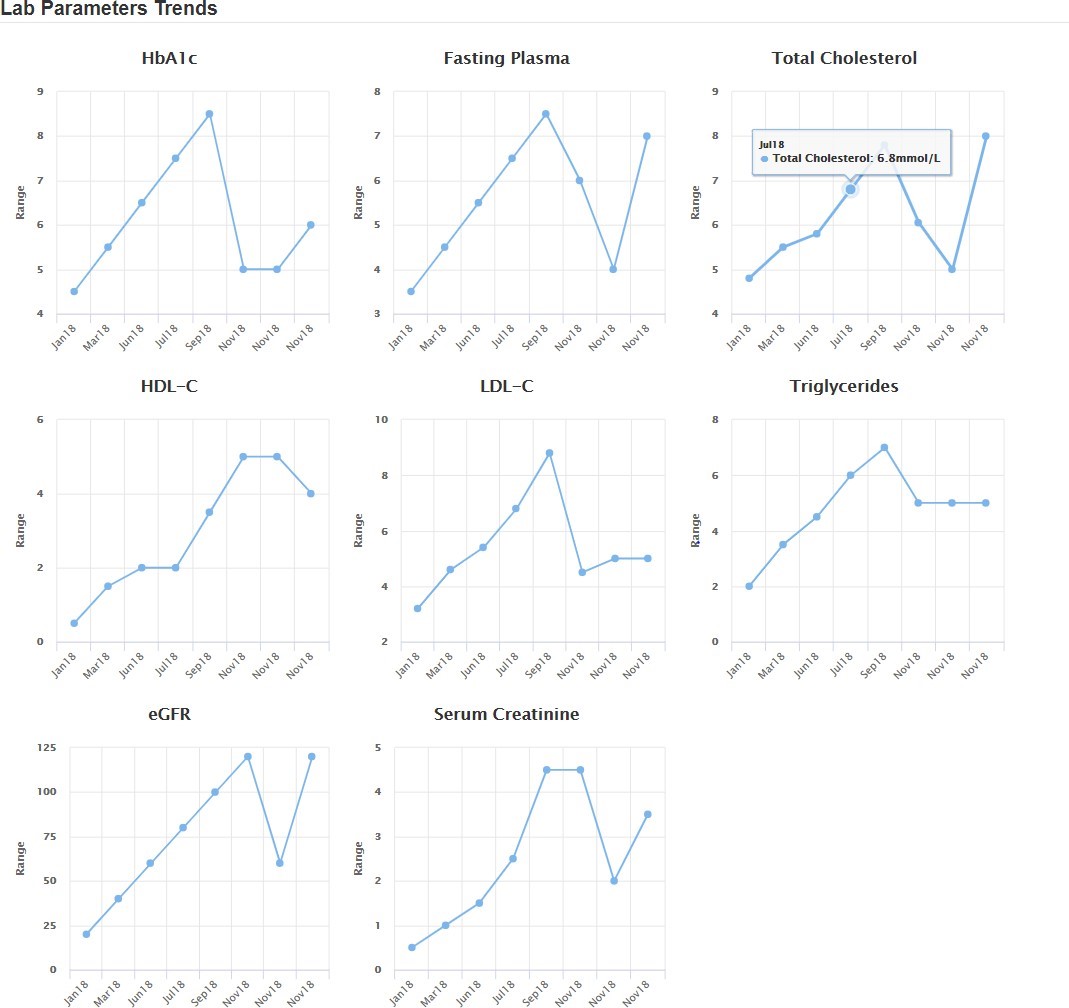}
 \caption{High chart line graphs}
 \label{fig:my_label}
\end{figure}

\section{Limitations}

The implemented app is working as expected, but there are some limitations such as the developed app can only display values of lab measures or clinical measures in mmol/L where diabetic measures can be also measured in mg/dl. When we decided to go with the gauge charts to display the measure ranges in color coding for all the three glycemic, renal, and lipid profiles, mg/dl will make the charts too big and complex, so we opted for another unit of measurement which made charts simple, helped in better visualization. This might be a problem for some physicians who are used to view values in mg/dl. We did not evaluate the implemented system. In the future, we will conduct a user study whose feedback will be used to improve the design and usability of the dashboard.

\section{Future Status}

Currently, the implemented app has dashboard functionalities, but in future, there is an idea to implement decision support capabilities, which will make the existing dashboard a fully functional Clinical Decision Support System. The decision support system will have a modules for alerting the provider. With the alert functionality it will be easier to know when the test is due. Diabetic complications are not captured at present. In the future, a system could be developed to monitor the complications. Along with these functionalities, implementation of Clinical Practice Guidelines will help physicians to practice Evidence-Based Medicine for making better decisions while designing a therapy. In the future, it might also be useful to automatically customize the dashboard on patient persona \cite{holden2017know}.

\section{Conclusion}
We have successfully developed a working diabetes dashboard and implemented it in OpenMRS. Development of a dashboard to summarize the information of individuals in a single screen, accessed with a single click within the patient health record page is accomplished. We hope the implemented system will help physicians to access the diabetic individual information all at a single place instead of going through multiple screens and a lot of clicks to see the results of patients. The implemented app saves physician time because of its design to show all the information at a single place. With this add-on feature to OpenMRS, it gives physicians more time to focus on the patient, rather than browsing multiple screens. This dashboard might be more beneficial if implemented in an outpatient setting. In future, the alert module will make the developed app as a decision support system which can help physicians in making clinical decisions.

\bibliographystyle{unsrt}
\bibliography{diabetes-bibliography.bib}  

\end{document}